# Pervasive Technology-Enabled Care and Support for People with Dementia: The State of Art and Research Issues

[1]**Sayan Kumar Ray**, [2]**Geri Harris**, [3]**Akbar Hossain**, [1]**NZ Jhanjhi**

[1]School of Computer Science, SCS Taylor's University, Subang Jaya 47500, Malaysia

[2]Auckland University of Technology, New Zealand

[3] Eastern Institute of Technology, New Zealand

Sayan.ray@taylors.edu.my ; Noorzaman.jhanjhi@taylors.edu.my ; Ahossain@eit.ac.nz

geri.harris@aut.ac.nz

***Abstract*—Dementia is a mental illness that people live with all across the world. No-one is immune; nothing can predict its onset. The true story of dementia remains unknown globally, partly due to denial of dementia symptoms and partly due to the social stigma attached with the disease. In recent years, dementia as a mental illness has received a lot of attention from the scientific community and healthcare providers. This paper presents a state-of-art survey of pervasive technology-enabled care and support for people suffering from Alzheimer's dementia. We identify three areas of pervasive technology support for dementia patients, focusing on: care, wellness and active living. A critical analysis of existing research is presented here exploring how pervasive computing, artificial intelligence (AI) and the Internet of Things (IoT) are already supporting and providing comforts to dementia patients, particularly those living alone in the community. The work discusses key challenges and limitations of technology-enabled support owing to reasons like lack of accessibility, availability, usability and affordability of technology, limited holistic care approach, and lack of education and information. Future research directions focusing on how pervasive and connected healthcare can better support the well-being and mental health impacts of Alzheimer's dementia are also highlighted.**

■ **THE INTRODUCTION** Dementia is a mental disorder affecting an estimated 50 million people across the globe; a figure that is expected to reach 152 million by 2050, accounting to 22% of the world's population (WHO, 2020). Dementia is a growing silent epidemic (a term coined in the early 1980s) and it is an important community issue. The word dementia means "without mind" [1]. It is a neurological disease that leads a person to lose the critical facilities that allow them to function in everyday life [2]. Dementia is a terminal illness, with no cure. It is not just an old person's disease, but a huge challenge for the whole population. People can develop it younger than 65 years old, no-one is immune, it strikes arbitrarily. The World Health Organisation (WHO) before COVID said that dementia is a global health priority. With an aging global population and a growing number of co-morbid diseases, the numbers living with dementia are expected to grow in the coming years. Alzheimer's Dementia (AD) accounts for 60-70% of all dementia cases (WHO, 2020) and is the focus of this review paper. Care homes and hospitals already treat a massive volume of patients with AD and with the increase in number they may reach the threshold pretty soon. Research, hence, shows that the home is the best place to stay for someone living with AD since it is the least distressing environment [2,3]. However, a high level of care and attention is needed to allow people to live with AD and feel safe in their own homes and communities. Technology-enabled assisted living has a crucial role to play in connecting these people to their whanau, friends, neighbours and voluntary community groups whom they are heavily reliant on for everyday support. According to A.J. Anstell et al. "Research applying technology to dementia has only recently gained mainstream attention." [4]. The cost of care for dementia globally is estimated at $305 billion annually [5], however the true cost of informal support certainly is unmeasurable.

In context to the discussions above, this paper presents a state-of-art survey of pervasive technology-enabled care and support for people suffering from Alzheimer's dementia. The motivation for this survey primarily centres around two research questions. Firstly, to synthesise state-of-the-art applications of pervasive technology to support people living with AD, and secondly, to define critical gaps in current scholarly and clinical knowledge [6-8].

The findings from this review can contribute to scholarly understanding and also to society how to better support the needs of a vulnerable group of society. By doing so we raise awareness amongst healthcare professionals about the potential for pervasive technology to support not just physical care but also emotional wellness and active living amongst persons with AD. Specific objectives from the review are to:

1. Recommend how community-based networks of informal caregivers can harness the potential of pervasive technologies to provide support to elderly patients along with formal caregivers.
2. Understand ways in which pervasive technology can be used to support mental well-being and enable elderly people with AD to continue living fulfilled lives in their own homes.

Achieving these objectives will produce ideas about how to build on the ability of carers to better respond to the day-to-day needs of AD sufferers; ideas which acknowledge that carers must try to predict needs that are constantly changing.

The rest of the paper is organised as follows. Firstly, we analyse the state of art in understanding pervasive technology-enabled support for dementia. Subsequently, we analyse and discuss the limitations of existing support, revealing critical gaps. Finally, conclusions about future research directions are drawn, with the proposal of innovative solutions that can broaden and integrate existing technology-based AD support [9-10].

■ **STATE OF ART** Existing research on technology-enabled care for mental health issues in older adults arising from AD reveals two primary themes: (a) use of pervasive technology solutions in private home-based care, and (b) integrating standalone technology solutions across patients, whanau, caregivers, and healthcare professionals, to provide holistic support.

Limited use of pervasive technology solutions in private home-based care
Understanding how technology can support people living with dementia is challenging since the symptoms of AD vary from person to person, manifesting as slow decline in memory, thinking and reasoning skills as depicted in Figure 1.

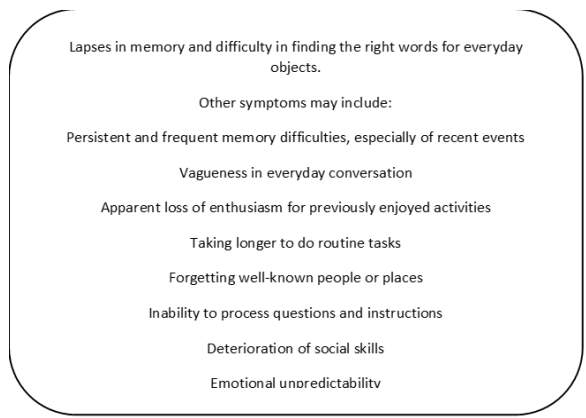

Figure 1. Symptoms of AD

Furthermore, what is known about how technology can help with AD is further complicated by the fact that AD is a gradual process that "typically progresses slowly in three general stages: early, middle and late (sometimes referred to as mild, moderate and severe in a medical context)" [5]. In a documentary by a New Zealand organisation - The Brains Trust - leading healthcare experts talk about making the world a more dementia-friendly place for sufferers. Sir Richard Faull, a neuroscientist with the Centre for Brain Research in NZ says we need to "support the person with the challenge in a way that gives them full value of their life, every day" [3-4]. Literature describes many technology-enabled solutions that currently aid in the management of daily activities for people living with AD, such as GPS tracking for wandering, fall-prediction software, scheduling for drug administration etc [11-12]. There is a paucity of understanding as to how pervasive technology can help support the *mental health* impacts of AD and the impacts of this type of dementia on the private home-based care of both the people with the disease and also their whanau and caregivers. Mental health impacts here refer to depression, anxiety, fear, and isolation. According to Berg-Weger and Morley (2020), referring to an article about loneliness in old age, "U.S.-based research suggests that between 17% - 57% of the world's population experience loneliness, a figure that increases for those who have mental and physical health concerns, particularly those with…depression, anxiety, and dementia" (M. Berg-Weger and J.E. Morley, 2020). Literature reveals that in supporting AD patients, technology "adoption rates remain low, in part due to lack of awareness or challenges in accessibility (including financial) and support".

A need to integrate standalone technology solutions

Extensive research exists about the use of standalone technologies to gather data in people's normal daily settings (e.g. to predict, detect, manage AD) (EDoN, 2020)[i]. Research has been conducted on how technology can help in predicting and diagnosing AD early based on behavioural patterns of patients. Furthermore, technology use for monitoring, care provision and maintaining routine activity in people living with AD is well understood. How to integrate the many technology solutions available in a way that provides adequate support to sufferers to continue to live meaningful, fulfilled lives in their own homes, and how technology can provide support in the mental well-being of these people, are research problem domains that are still nascent in nature.

According to recent literature, technology support is not well integrated across patients, whanau, caregivers, and healthcare professionals, to provide holistic support. "Professionals and society…seem to lack an applied understanding of the potential of assistive technology in dementia because it is not being integrated into mainstream dementia care practice." There are calls for future technology-enabled support for dementia to enable seamless integration of technology and human and integrating care with lifestyle [13].

■ **EXISTING TECHNOLOGY-BASED SUPPORT**

Surveying existing literature revealed technology-enabled support for AD in a number of discrete but interrelated areas: (i) physical care, (ii) well-being and (iii) active living. Discussion of each area of support is addressed in this section, using a consistent format whereby we define the primary needs for support in each area and then critically review current technology-based support for that aspect of living with AD. We summarise primary needs within each support area, map technology solutions to the needs they support and identify limitations of current application of pervasive technology across the three areas of AD support.

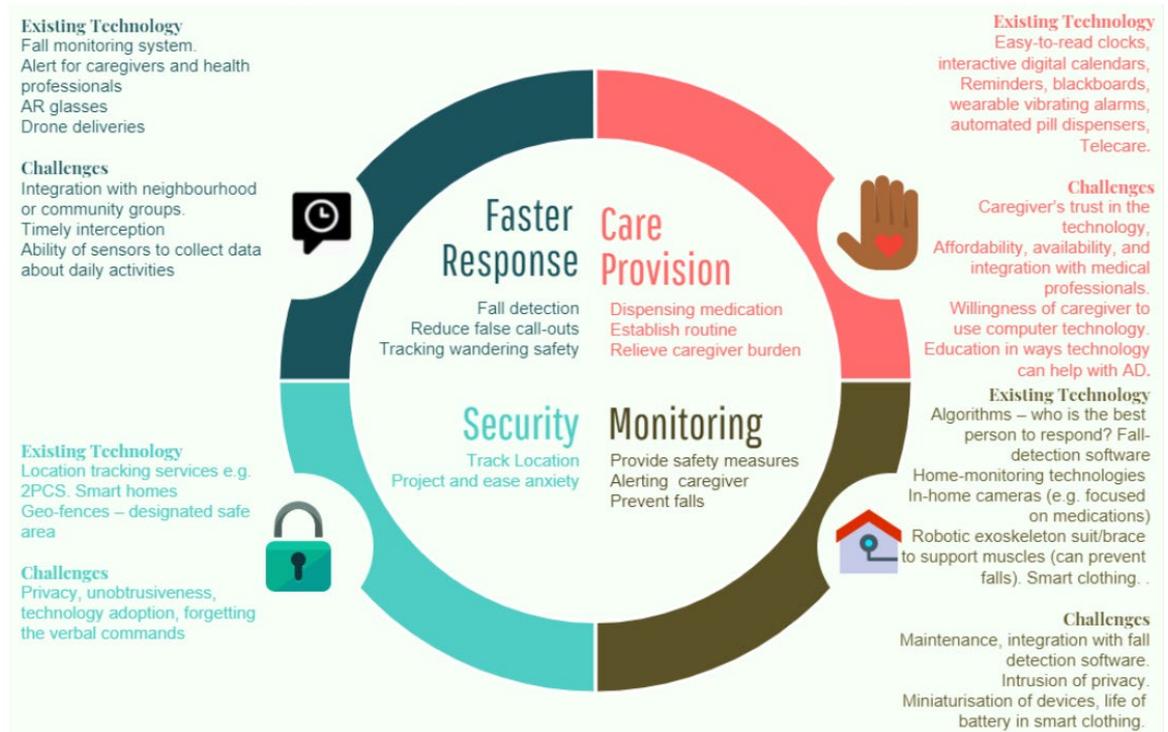

Figure 2: **Attributes of care and technology solutions**

Technology-based physical care
The cost of care-facilities is a driver for extended home-based care for AD sufferers. However, caring for someone living with AD relies heavily on community-based support options. Carers avail of technology-enabled solutions to assist them in supporting the physical everyday needs of the person living with dementia, for example, fall detection apps that alert a nearby carer to respond in the case of a fall. Current care provision solutions based on technology assist with dispensing medication, telecare, webcams, fall detection, safety monitoring, alerts for taxi to come, in-home safety devices, Smartphone apps, wearable detection devices (fobs), assistive technology, lifestyle monitoring technology. These currently assist people living with AD by improving physical safety, establishing routines, easing anxiety, maintaining independence and contributing to physical wellness. Technology contributes to keeping the person safe when their caregiver isn't physically with them. wireless technology such as GPS tracking is helping to manage patient wandering by using real-time notifications of a patient's whereabouts. Pervasive technologies that embed microchips in everyday things (e.g. a kettle) can communicate information to help patients and caregivers to monitor symptoms and activity. For example, an alert can be sent to a caregiver if a patient's kettle is left switched on and could pose a danger. Self-reporting and data collected from sensors round the clock (24x7) can be analysed by healthcare professionals to ensure that appropriate care is being given. Artificial Intelligence is helping patients with their daily routines, for example cooking, through the use of intelligent-assistive technology [14]. Figure 1 contains an overview of key aspects of physical care for AD sufferers along with a synthesis of current technologies supporting caregivers. In the figure, the major limitations affecting the use of these technology-enabled support mechanisms are identified.

Different technology-based products offer a range of care features including reminder notifications/beeps to alert the person to take their medication, vibration alerts, and calendar reminders. New technology is beginning to allow caregivers to sync apps to enable even better support, provide greater peace-of-mind for the patient's family, and allay some of the angst of carers at shouldering the responsibility of care. However, the challenge remains as to how to integrate the various standalone technologies for physical care of people living with AD.

Technology to support wellness
Technology that assists a person living with AD to look after their emotional needs is different to technology supporting aspects of physical care. Loneliness and social isolation have been shown to significantly impact older adults suffering from AD. Mental health problems are under-identified by health-care professionals and older people themselves, and the stigma surrounding these conditions makes vulnerable people reluctant to seek help or ask for immediate support even from near and dear ones or from the community [15]. Whilst AD is becoming more 'public' and society is being educated that sufferers are just normal people who we shouldn't be scared of, for the sufferers evidence shows growing anger, alcohol misuse, fear and resentment [16-17]. According to The Brains Trust, "knowing what is going on in your mind is very important for your peace-of-mind". Treating AD sufferers with humanity, love and respect should be at the forefront of care provision, though this care requires supporting their emotional wellness in addition to their physical care as discussed in an earlier section. Some commonly experienced mental health impacts of AD affecting wellness include loneliness, focus on somatic symptoms, difficulty communicating or expressing feelings, diminishing memory abilities, loss of dignity, rapid mood changes or overreacting, anxiety/fear, loss of a sense of familiarity, difficulty feeling love, loss of a sense of identity, stigma and depression [18]. Technological advances aimed at reducing these mental health issues are receiving increased attention from scientists and researchers. Devices that increase attachment and interest in activities, improve comfort, promote fun and play, encourage participation and social inclusion, and even provide companionship are becoming more prevalent in the care of AD. These devices make use of movement, noise and visual stimulation to improve cognitive function by allowing users to communicate feelings using pictures, icons, and symbols, for instance. Or to recall memories of how an event made them feel. Robotic pets, for example, offer low maintenance companion pets targeted at reducing loneliness and easing anxiety. Some of the more advanced robotic pets such as the Joy For All pet use sensors to respond to human attention like petting, providing a realistic experience for the patient. Other electronic devices designed with large buttons and easy to use interfaces are aimed at making it easy for users to connect to friends, Whanau and the community. Pre-programmed functionality can motivate engagement with these devices in an intuitive and straightforward way for users. In using these products, questions exist about what constitutes a medical intervention versus private or informal care. Should the products be available through healthcare professionals or remain the responsibility of the patient's family to provide?

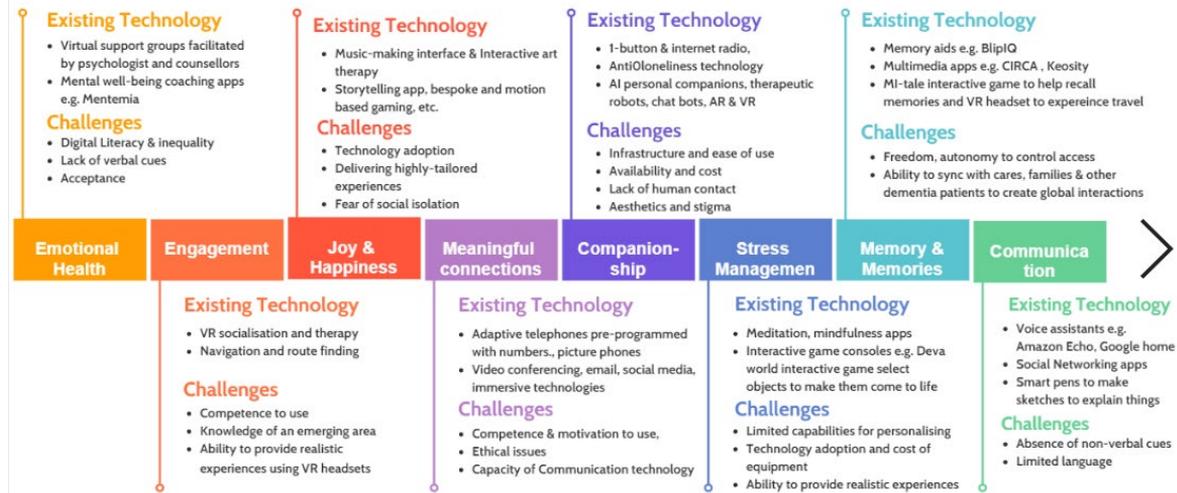

Figure 3: Attributes of wellness and technology solutions

New Zealander Sir John Kirwan launched a mental well-being coaching app in 2018, based on clinical advice and best practice in mental health. Although not specifically produced for dementia patients, Mentemia aims to support users to "Feel good. Bounce back from stress. Sleep better."

Figure X contains a synthesis of the main technologies currently targeted at reducing the mental health impacts of AD in a number of areas, together with an outline of the main limitations of these technologies.

Active living through pervasive technology
People living with AD can experience 'normal' living via pervasive technology that helps them live a healthy lifestyle, with independence and agency, while free of stigma. Emerging technologies provide a positive impact on AD patients by helping people to keep doing all the things that they love to do. Ambient assisted living technologies aim to see the person, not the condition and support the person to feel comfortable, have a familiar sense of safety, and a feeling of being at home [19-21].

■ IMPORTANT GAPS IN EXTANT LITERATURE
The literature analysed in the preceding sections illustrates that assistive technology for the physical care needs of people living with AD has received much attention from health practitioners and academics. Less is known, however, about applying pervasive technologies to enhance mental wellness through giving pleasure, or in facilitating meaningful, active living for older adults living with AD. It is important that future research work considers multiple perspectives (such as the sufferer, their whanau, caregivers, healthcare professionals). Due to space limitations and in the interests of attending to an emerging technology area, the ensuing discussion focuses on identifying important challenges and limitations of existing technology-based support primarily from the perspective of the person living with AD. Challenges with current applications of technology to provide care, improve wellness and facilitate active living can be broadly organised into categories.

Accessibility

| |
|---|
| Problems of technology literacy, user-friendliness, cost, availability, eligibility |
| Lack of education and information – not knowing what is available, competence, skills, information equity |

Holistic care

| |
|---|
| Lack of integration between technology and caregivers – to give holistic care |
| Training required for caregivers in how to use the assistive-technologies alongside in-person care |

National healthcare perspective

| |
|---|
| Tendency for medicinal solutions over technology adoption – due to lack of trust in technology, reluctant adoption of assistive-technologies over human contact |
| Variable priority of giving people access to assistive-technology (at a national policy level) |

- Preference for human contact in care provision
- Stage of dementia
- Difficulties arise using Smartphones and apps with only minor cognitive limitations. Could make use unreliable (e.g. if for taking medication)
- Ethical considerations – privacy, security, intrusiveness, dignity
- Current advances in technology-enabled support require specialised clinical environment for testing – limited research done in home-based care settings

Focus on wellness and active-living in this section…

Wellness-specific limitations

Some of the important challenges to overcome in supporting emotional wellness through assistive technology include:

- Ways in which technology can provide more engagement, joy, happiness, delight, enhanced meaningful connections, companionship, soothing, calmness, create memories.
- Preserving dignity
- Maintaining sense of identity and expression of personality
- Ensuring that the technology used does not reduce the person's opportunity for activity (e.g. technological clothing designed to prevent undressing at inappropriate times can restrict movement for some activities)
- Identifying the actual well-being needs of dementia patients that technology can solve.

Active living-specific limitations

Some of the important challenges to overcome in supporting active living through assistive technology include:

| |
|---|
| How technology can assist living alone for as long as possible |
| Knowing first-hand information of how aged care facilities are managing elderly people with dementia |
| Calls for testing, scan for a predisposition, forecast what the future holds |
| Tailored experiences and personalisation are limited. Elderly people living with dementia are seeking freedom and autonomy in how technology supports them to lead meaningful active lives |

Attributes of future technology-enabled support for AD

| |
|---|
| Teach robots to recognize "nonverbal human interactions on their own and respond to the behaviors accordingly" (Live In Home Care, n.d.) |
| "develop more advanced and responsive humanoid robots to help the elderly" (Live In Home Care, n.d.) |
| Overcome resistance to new things |
| Ability to tailor technology to the person's needs |
| Simplify |
| Technology can be overwhelming |
| Humanise |
| Heterogeneous technology |
| Seamless integration of technology and human |
| Integrate care with lifestyle |

"The pace of technology development requires urgent policy, funding and practice change, away from a narrow medical approach to a holistic model that facilitates future risk reduction and prevention strategies, enables earlier detection and supports implementation at scale for a meaningful and fulfilling life with dementia" [22-23]

# ■ CONCLUSION

Future innovative solutions to support mental wellness and integration of care are presented in this section, along with future potential research areas. The work discusses key challenges and limitations of technology-enabled support owing to reasons like lack of accessibility, availability, usability and affordability of technology, limited holistic care approach, and lack of education and information. Future research directions focusing on how pervasive and connected healthcare can better support the well-being and mental health impacts of Alzheimer's dementia are also highlighted.

# ■ REFERENCES

1. P. Sachdev, Is It Time to Retire the Term "Dementia"?, *J. Neuropsychiatry*, 2000
2. C. Meng-Yee, Dementia: The Brains Trust, Episode 1- Deborah and Anne Pead, New Zealand Herald [Online] Available: https://www.nzherald.co.nz/nz/dementia-the-brains-trust-episode-1-deborah-and-anne-pead/7F5U3KYYXY4XRU2YMEL7O4WOIE/
3. C. Meng-Yee, Dementia: The Brains Trust, Episode 2 - Warwick and Pummy Hickling, New Zealand Herald [Online] Available: https://www.nzherald.co.nz/nz/dementia-the-brains-trust-episode-2-warwick-and-pummy-hickling/FYALP775CGYKLQG4WJWMKVA4GM/
4. A.J. Anstell, B.N. Hoey, L.A. Mihailidis, N.C. Robillard, Technology and Dementia: The Future is Now, J. Dementia and Geriatric Cognitive Disorders, Volume 47, Issue 3:131-139, (2019)
5. W. Wong, Economic Burden of Alzheimer Disease and Managed Care Considerations, Am. J. Managed Care, Volume 26, Issue 8, 2020
6. Alzheimer's Association, Stages of Alzheimer's, [Online] Available: https://www.alz.org/alzheimers-dementia/stages, 2021
7. Ray, S. K., Pawlikowski, K., & Sirisena, H. (2009). A fast MAC-layer handover for an IEEE 802.16 e-based WMAN. In AccessNets: Third International Conference on Access Networks, AccessNets 2008, Las Vegas, NV, USA, October 15-17, 2008. Revised Papers 3 (pp. 102-117). Springer Berlin Heidelberg.

Department Head8    IT professional

8. Shahid, H., Ashraf, H., Javed, H., Humayun, M., Jhanjhi, N. Z., & AlZain, M. A. (2021). Energy optimised security against wormhole attack in iot-based wireless sensor networks. Comput. Mater. Contin, 68(2), 1967-81.
9. Singhal, V., Jain, S. S., Anand, D., Singh, A., Verma, S., Rodrigues, J. J., ... & Iwendi, C. (2020). Artificial intelligence enabled road vehicle-train collision risk assessment framework for unmanned railway level crossings. IEEE Access, 8, 113790-113806.
10. Humayun, M., Ashfaq, F., Jhanjhi, N. Z., & Alsadun, M. K. (2022). Traffic management: Multi-scale vehicle detection in varying weather conditions using yolov4 and spatial pyramid pooling network. Electronics, 11(17), 2748.
11. Lim, M., Abdullah, A., Jhanjhi, N. Z., Khan, M. K., & Supramaniam, M. (2019). Link prediction in time-evolving criminal network with deep reinforcement learning technique. IEEE Access, 7, 184797-184807.
12. Haslam-Larmer, L., Shum, L., Chu, C. H., McGilton, K., McArthur, C., Flint, A. J., ... & Iaboni, A. (2022). Real-time location systems technology in the care of older adults with cognitive impairment living in residential care: A scoping review. Frontiers in psychiatry, 13, 1038008.
13. Vuong, N. K., Chan, S., & Lau, C. T. (2015). mHealth sensors, techniques, and applications for managing wandering behavior of people with dementia: A review. Mobile Health: A Technology Road Map, 11-42.
14. Kearns, W. D., Fozard, J. L., & Nams, V. O. (2016). Movement path tortuosity in free ambulation: relationships to age and brain disease. IEEE journal of biomedical and health informatics, 21(2), 539-548.
15. F. Meiland, Technologies to Support Community-Dwelling Persons With Dementia: A Position Paper on Issues Regarding Development, Usability, Effectiveness and Cost-Effectiveness, Deployment, and Ethics, *JMIR Rehabilitation Assistive Technology*, 4(1): e1, 2017
16. L. Ault, R. Goubran, B. Wallace, H. Lowdon and F. Noefel, Smart home technology solution for night-time wandering in persons with dementia, J. Rehabilitation and Assistive Technologies Engineering, SAGE Publications, 2020
17. Live In Home Care, Age-Technology-How AI and IoT in Healthcare Can Help Dementia Patients, [online] Available: https://www.liveinhomecare.com/age-technology-how-ai-and-iot-in-healthcare-can-help-dementia-patients-2/age-technology (n.d.)
18. M. Berg-Weger and J.E. Morley, Loneliness in Old Age: An Unaddressed Health Problem, J. Nutrition, Health & Aging, Volume 24:243–245, 2020
19. Handley, M., Bunn, F., & Goodman, C. (2017). Dementia-friendly interventions to improve the care of people living with dementia admitted to hospitals: a realist review. BMJ open, 7(7), e015257.
20. World Health Organisation (WHO), Dementia fact sheet, [online] Available: https://www.who.int/news-room/fact-sheets/detail/dementia
21. World Health Organisation, Coronavirus disease 2019 (COVID-19) situation report-79. World Health Organisation, [Online] Available: http://www.who.int/docs/default-source/coronavirus/situation-report/202000498-sitrep-79-covid-19.pdf?sfvrsn=4796b1443_4, 2020
22. Taj, I., & Zaman, N. (2022). Towards industrial revolution 5.0 and explainable artificial intelligence: Challenges and opportunities. International Journal of Computing and Digital Systems, 12(1), 295-320.
23. Borson, S., Frank, L., Bayley, P. J., Boustani, M., Dean, M., Lin, P. J., ... & Ashford, J. W. (2013). Improving dementia care: the role of screening and detection of cognitive impairment. Alzheimer's & Dementia, 9(2), 151-159.
24. Ray, S. K., Sinha, R., & Ray, S. K. (2015, June). A smartphone-based post-disaster management mechanism using WiFi tethering. In 2015 IEEE 10th conference on industrial electronics and applications (ICIEA) (pp. 966-971). IEEE.
25. Chaudhuri, A., & Ray, S. (2015). Antiproliferative activity of phytochemicals present in aerial parts aqueous extract of Ampelocissus latifolia (Roxb.) Planch. on apical meristem cells. Int. J. Pharm. Bio. Sci, 6, 99-108.
26. Ray, S. K., Pawlikowski, K., & Sirisena, H. (2009). A fast MAC-layer handover for an IEEE 802.16 e-based WMAN. In AccessNets: Third International Conference on Access Networks, AccessNets 2008, Las Vegas, NV, USA, October 15-17, 2008. Revised Papers 3 (pp. 102-117). Springer Berlin Heidelberg.
27. Hossain, M. A., Ray, S. K., & Lota, J. (2020). SmartDR: A device-to-device communication for post-disaster recovery. Journal of Network and Computer Applications, 171, 102813.

[i] Early Detection of Neurodegenerative Diseases (EDoN) 2020 project on wearable devices https://osf.io/u49z5